\documentclass[letterpaper, 10 pt, conference]{ieeeconf}  % Comment this line out if you need a4paper

\IEEEoverridecommandlockouts                              % This command is only needed if 
\usepackage{graphics} % for pdf, bitmapped graphics files
\usepackage{graphicx}
\usepackage{amsmath} % assumes amsmath package installed
\usepackage{amssymb}  % assumes amsmath package installed
\usepackage[font=footnotesize, labelsep=period]{caption}
\usepackage{color}
\usepackage[noadjust]{cite}

\usepackage{dblfloatfix}

\usepackage{balance}

\makeatletter
\let\NAT@parse\undefined
\makeatother
\usepackage{hyperref}

\newcommand{\rline}{{\mathbb R}}

\newcommand{\bbm}[1]{\left[\begin{matrix} #1 \end{matrix}\right]}

\title{
\LARGE 
\bf
Homomorphic Encryption-enabled Distance-based Distributed Formation Control with Distance Mismatch Estimators
}

\author{Mariano Perez Chaher$^{1}$, Bayu Jayawardhana$^{1}$, and Junsoo Kim$^{2}$% <-this % stops a space
    \thanks{ $^{1}$M. Perez Chaher and B. Jayawardhana are with the Engineering and Technology Institute Groningen, Faculty of Science and Engineering, University of Groningen, 9747AG Groningen, The Netherlands.}%
    \thanks{ $^{2}$J.~Kim is with the Division of Decision and Control System, KTH Royal Institute of Technology, Sweden.}
}

\begin{document}

\maketitle
\thispagestyle{empty}
\pagestyle{empty}

%%%%%%%%%%%%%%%%%%%%%%%%%%%%%%%%%%%%%%%%%%%%%%%%%%%%%%%%%%%%%%%%%%%%%%%%%%%%%%%%
\begin{abstract}
This paper considers the use of homomorphic encryption for the realisation of distributed formation control of multi-agent systems via edge computer. In our proposed framework, the distributed control computation in the edge computer uses only the encrypted data without the need for a reset mechanism that is commonly required to avoid error accumulation. Simulation results show that, despite the use of encrypted data on the controller and errors introduced by the quantization process prior to the encryption, the formation is able to converge to the desired shape. The proposed architecture offers insight on the mechanism for realising distributed control computation in an edge/cloud computer while preserving the privacy of local information coming from each agent.
\end{abstract}


%%%%%%%%%%%%%%%%%%%%%%%%%%%%%%%%%%%%%%%%%%%%%%%%%%%%%%%%%%%%%%%%%%%%%%%%%%%%%%%%

\section{Introduction}  \label{SEC:introduction}

Encryption standards such as the Advanced Encryption Standard (AES) have provided a reliable method to ensure data remains private \cite{AES_comp}. However, the development of network computing and the Internet of Things, that is, the collaboration of different nodes within a network, have brought up concerns on the privacy and security of data that require new solutions \cite{Security_IoT}. The introduction of 5G communication technology and beyond, which guarantees ultra low-latency and high reliability of the wireless network, paves way for the implementation of cloud-based or edge-based control systems to control complex systems over a wide geographical area with minimal infrastructure footprint. While real-time data can reliably be transferred over air without wired/cabled connections, the issue of privacy and safety of critical real-time data transmission has been a major concern in the adoption of 5G wireless network by the high-tech industry. 

Data sharing when working with different nodes in a network is inevitable. The security compromise is created by the availability of plaintext data in an untrustworthy third-party managed node that can be directly accessed by malicious agents. This problem manifests itself in %the area of control engineering in the form of 
Networked Control Systems (NCS) \cite{NCS_sec} when cloud-based or edge-based control systems are deployed. In order to perform the operations needed, an NCS using current encryption standards requires data to be decrypted at each node. The distribution of plaintext data across the network can be avoided by using Homomorphic Encryption (HE). By allowing for the manipulation of data in its ciphertext form it provides another layer of security \cite{HE_survey} which has been used in NCS to guarantee the privacy of real-time data of sensor and control signals \cite{alexandru2020,kim2020,darup2020,kogiso2020}.

Although HE provides further security to network computing by allowing for addition and multiplication of ciphertexts, it also has its own drawbacks depending on the chosen encryption algorithm.
For HE schemes based on the Learning With Errors (LWE) problem,
the limitations are the gradual corruption of data depending on the amount of ciphertext operations,
and the number of operations other than addition and multiplication, in case bootstrapping techniques are not utilized for the sake of real-time operation \cite{kim2020}.
%For the Learning with Errors (LWE) approach used for Fully Homomorphic Encryption (FHE), the limitations are the gradual corruption of data depending on the amount of ciphertext operations and the inability to encrypt decimals \cite{kim2020}.
These issues are especially challenging for dynamic controllers which require recursive updates of the system's state. Two methods exist to mitigate this limitation.

An approach proposed by Murguia et al. \cite{Reset_HE} is to periodically reset the controller's state to its initial value. Alternatively, bootstrapping can be used to refresh the state unlimitedly \cite{Boot_FHE}. These methods increase computational complexity of the process or lead to performance degradation, making them less attractive for practical uses. %In a recent work, Kim et al. \cite{Shim_2019,kim2020} introduce a new encryption approach in feedback control systems that allows to infinite update of the controller’s state without its reset, nor data corruption. In this paper, we extend the applicability of FHE as proposed in \cite{Shim_2019} to distributed control systems. In particular, we present a scheme that allows for the control of a distributed dynamic multi-agent formation using homomorphic encryption and the control of error growth on the system's encrypted state using the method developed in \cite{Shim_2019}. 
In this work, we consider integral controller design so that, following the observations of previous research \cite{Shim_2019,Integer_Dynamic}, it is made possible to update the encrypted system's state for an infinite time horizon without its reset, nor performance degradation. In particular, we present a scheme that allows for the control of a distributed dynamic multi-agent formation using homomorphic encryption.

Distributed formation control refers to the design of distributed controllers for steering all robots to reach and maintain a prescribed formation shape based on local information from on-board sensor systems. The desired formation shape is typically defined using the specific relative information to the neighbours, such as, distance, relative position, feature or bearing information. Correspondingly, the local controller uses the same information obtained from the sensor to steer the robot towards the right direction and distance. We refer to \cite{oh2015} for a comprehensive review on the gradient-based formation control approaches. Among these methods, distance-based approach has been widely adopted as it is based on local coordinate frame, as well as, local relative measurement, enabling full distributed implementation of the control law \cite{marina2015}. Once the formation shape can be maintained, high-level group tasks can be carried out accordingly, such as, performing group motion \cite{de2016distributed}.      
 
%The encrypted formation controller we work towards makes use of existing control laws that determine an agent's movement to reach a position within a formation. Different methods exist to calculate the desired position, each of which with its benefits and drawbacks. Distance-based controllers simplify the formation control problem by requiring only knowledge of the inter-agent distances \cite{oh2015}, and by extension, reducing computational power needed to perform HE. Furthermore, this method allows agents to rely only on local information, which creates the possibility to use distributed control.

%Distributed systems have been of interest in computing applications due to their potential for scalability, agent autonomy and distribution of computing load. In this work, the considered approach to distributed formation control makes use of the rigid properties of certain formation distributions to obtain stability \cite{marina2015}. 

While a distance-based formation control law can be implemented locally, it can induce undesired formation shapes and group motion when there are mismatches in the distance constraints or measurement biases between pairing agents \cite{mou2016}. In order to tackle this unwanted behaviour, dynamic estimators can be deployed which compensate these mismatches or biases \cite{marina2015}. 

In this paper, we implement and evaluate the use of HE-enabled distance-based formation control law with mismatch estimators via edge/cloud computer. 
%However, one of the challenges of using only local information is the potential disagreement between agents \cite{mou2016}. To account for the undesired motion caused by these mismatches in information, we implement a dynamic controller capable of stabilizing the formation by estimating disagreements \cite{de2016distributed}. 
%The final suggested encrypted formation is a potential candidate for edge-based or cloud-based control. 
%The computation of the formation control in the cloud is still done distributedly, in the sense, that 
The control computation for each agent is performed independently of that of the other agents; hence it is ``locally'' computed in the edge/cloud. This allows us to offload computational resources from the robotic agent to the edge/cloud and enable complex  tasks computation (such as, SLAM and collision avoidance) to be done in the same local environment in the edge/cloud. In practice, the real-time transmission of the sensor data and control input data will be enabled by means of low-latency wireless network of 5G or beyond. The use of infinite-time horizon HE in this context is to secure the privacy of both the sensor data and control data in the edge/cloud while enabling us to achieve the desired formation shape. We propose the use of scaled logarithmic  quantizer prior to the encryption for maintaining precision and investigate the role of encryption key length to the overall performance. In our simulation, we show that the formation performance of the multi-agent systems does not deteriorate in spite of the presence of mismatches.  
 %  By making use of the low latency connections of a 5G network, an agent with the limited capability to sense distances to its neighbours could be monitored and managed from an edge gateway. By offloading computing power from the robotic agent, the costs for formation scalability are reduced, while maintaining security through the use of HE.

The paper is organized as follows. Section \ref{SEC:preliminaries} presents the preliminaries on the  encryption process and the distributed formation control with estimators. In Section \ref{SEC:designed_system}, we present the use of scaled logarithmic quantizer and sets out the HE-enabled distributed formation control architecture. % proposes the required system that allows for the operation of the robot control, 
We present the simulation setup and results in Section \ref{SEC:simu_set}. The conclusions and future works are presented in Section \ref{SEC:conclusions}. % describes future improvements and considerations on the application of HE, and Section \ref{SEC:conc} outlines the conclusions of the paper.

\section{Homomorphic Encryption and Distance-based Distributed Formation Control with Estimator
%FULLY HOMOMORPHIC ENCRYPTION AND DISTANCE-BASED DISTRIBUTED FORMATION CONTROL WITH ESTIMATOR
} \label{SEC:preliminaries}

%Given that this paper consists on the combination of two research topics, the following preliminaries are divided into four sections. 
In this section, we will provide some preliminaries that covers the two main topics on HE and on robust distributed formation control which employs distributed state estimators to compensate for distance mismatches and measurement biases. In the first two subsections, we present %The first two cover the mathematics of 
HE for standard feedback control systems \cite{kim2020} and a method to ensure the encryption in the closed-loop systems is operational for an infinite time horizon without the need to reset the states 
%dynamic controller can be encrypted for an infinite time horizon from 
\cite{Shim_2019}. In the following two subsections, we introduce the distributed formation control methods  \cite{de2016distributed} with its associated dynamic estimator that compensates for the measurement bias. %the dynamic estimation process that provides the final control law to be encrypted.

\subsection{Homomorphic Encryption} \label{SEC:FHE_pre}

%To fulfill the requirements of FHE both addition and multiplication should be capable within the ciphertext space. This section will introduce the LWE-based method used to do so.

%Let us revisit FHE that allows for both the addition and multiplication operations to be done in ciphertexts. Firstly, we need to define a few notations. % Let us introduce some notation. 
%{\color{red} The plaintext space $\mathcal P$ is the space of integer sequence $\{z_i\}_{i=1,\ldots,n}$, $z_i\in\mathbb Z$.} %An integer in the plaintext space is in the set $[p]$ denoted as $\mathcal{P}$. 
%An element $m\in \mathcal{P} $ represents a message in its plaintext form. {\color{red} For a given $m\in \mathcal{P}$, let $p$ be chosen as a power of $10$ such that $|m| < p/2$ with $|\cdot|$ be the $l_{\infty}$ norm.} Let $L$ be a power of $10$ so that $\mathbb{Z}_q$ is a set of integers with modulo $q$, where $q = Lp$. Finally a secret key used to decrypt and encrypt a message $m$ is denoted by $sk$, which is an integer vector of size $N$ such that $sk \in \mathbb{Z}^{N}_{q}$.

%{\color{green}

%($@$ Mariano, I just revised the above paragraph as the below (blue), where I don't erase the original one because I actually could not understand the red sentences above, which are not reflected in the below.
%Please check whether the blue one itself is fine, or correct the revised one. 
%)
%}

{
Let us revisit HE that allows for both the addition and multiplication operations to be done in ciphertexts. Firstly, we need to define some notation.
With a positive integer $p\in\mathbb N$,
which is chosen as a power of $10$ for convenience,
we define the space of plaintexts, as $\mathbb Z_p:= \{m\in\mathbb Z: -\frac{p}{2}\le m < \frac{p}{2}\}$,
whose cardinality is $p$,
so that an element $m\in \mathbb Z_p $ represents a message in its plaintext form. Let $L$ be also a power of $10$ and $q= Lp$, so that
the space $\mathbb Z_q$ of ciphertexts is defined as $\mathbb Z_q:=\{m\in\mathbb Z: -\frac{q}{2}\le m < \frac{q}{2}\}$.
Finally, a secret key used to decrypt and encrypt a message $m$ is denoted by $sk$, which is an integer vector of size $N$ such that $sk \in \mathbb{Z}^{N}_{q}$.
}

When a message $m \in \mathbb{Z}^{n\times 1}$ is encrypted, two new random matrices are generated: $A \in \mathbb{Z}^{n \times N}_{q}$, and injected error $e \in \mathbb Z^n$ whose elements are sampled from zero-mean discrete Gaussian distribution\footnote{For the error distribution, we assume that the parameter $L$ is chosen sufficiently large and neglect the probability that $\left|e \right|\ge r/2$, with some $r<L$. }, where $n$ is the dimension of $m$. Using these components, the message $m$ is encrypted as follows
\begin{equation} 
\label{eq:Enc_func}
\text{Enc}(m) = [(-A\cdot sk + Lm + e) \bmod q, A] =: %\rightarrow 
\textbf{m},
\end{equation}
where
$\mathrm{mod}$
is the (component-wise) modulo operation defined as
$a\bmod q:= a - \lfloor (a +(q/2)) / q \rfloor q$
for
$a\in\mathbb Z$ and $q\in\mathbb N$, $\textbf{m} \in \mathbb{Z}_{q}^{n\times(N+1)}$ is the encrypted form of $m$, $[\cdot, \cdot]$ denotes the concatenation of two matrices, and $\lfloor\cdot\rfloor$ the floor function.

%The natural step after encryption is 
Subsequently, the decryption process can be done by using a secret key vector $s:=[1,sk^T]^T$. Hence, the ciphertext $\textbf{m}$ is decrypted as
\begin{equation} 
\text{Dec}(\mathbf{m}) = \bigg\lceil \frac{(\mathbf{m} \cdot s)\bmod q}{L} \bigg\rfloor = \bigg\lceil\frac{Lm+e}{L}\bigg\rfloor =: m,
\label{eq:Dec_func}
\end{equation}
where $\lceil\cdot\rfloor$ is the element-wise rounding operation.

These particular encryption and decryption processes allow us to use homomorphic operations between plaintexts and ciphertexts. In other words, the following arithmetic holds true:
\begin{equation}
\begin{aligned}
\label{eq:Dec_simp}
\text{Dec}(\text{Enc}(m_1) + \text{Enc}(m_2)\bmod q) = m_1 + m_2,
\end{aligned}
\end{equation}
as long as $\left|m_1+m_2\right|<p/2$.

Regarding the multiplication $m_1 \times m_2$ over encrypted data,
we let the multiplicand $m_2\in \mathbb Z_p$ be encrypted with $\text{Enc}$,
and
use a different encryption process
for the multiplier  $m_1\in\mathbb Z_p$,
denoted by $\text{Enc2}$,
%However, homomorphic multiplication requires a different encryption process to calculate $m_1m_2$ with multiplier $m_1$ and multiplicand $m_2$.
%Whilst $m_2$ can be encrypted using the previously shown function $\text{Enc}(\cdot)$, the multiplier $m_1$ must be encrypted as follows
defined as
\begin{equation} 
\text{Enc2}(m_1) = m_1 R + \mathbf{O} =:\mathbf{M_1},
\label{eq:HE_add1}
\end{equation}
where $\mathbf{O}=\text{Enc}(0_{\log(q)-(N+1)\times 1})$ is an encrypted zero vector, and $R$ is defined as:
\begin{equation} 
R:=[10^0, 10^1, 10^2,\ \cdots, 10^{\log(q)-1}]^T \otimes I_{N+1},
\label{eq:HE_add2}
\end{equation}
with $\otimes$ being the Kronecker product. %, identity matrix $I$.

Then, as any ciphertext $c$ in $\mathbb{Z}^{1\times (N+1)}_{q}$ can be represented as $c = \sum^{\log(q)-1}_{i=0}10^i\,c_i$, where $c_i$'s components are single digits from 0 to 9, a decomposition function $D$ can be used to decompose the ciphertext by its digits in the form
\begin{equation} 
D(c):=
\begin{bmatrix}
   c_0, c_1,\ \cdots, c_{\log(q-1)}
\end{bmatrix}
.
\label{eq:HE_add3}
\end{equation}

Since $c=D(c)R$, the multiplication of ciphertexts can be computed as
\begin{equation} 
\mathbf{M_1} \times_\mathcal{C} \text{Enc}(m_2) := D(\mathbf{m_2}) \cdot \mathbf{M_1}\bmod q.
\label{eq:HE_mult_1}
\end{equation}
Consequently, we have the following property of homomorphic multiplication using both $\text{Enc}$ and $\text{Enc2}$ %may be described as
\begin{equation} 
\begin{aligned}
\text{Dec}\Big(\text{Enc2}(m_1) \times_\mathcal{C} \text{Enc}(m_2)\Big) = m_1m_2,
\label{eq:b}
\end{aligned}
\end{equation}
as long as $\left| m_1 m_2\right|<p/2$.

Instead of using two encryption processes as above for multiplication operation, one can also consider the use of LWE-based cryptosystem in \eqref{eq:Enc_func} and \eqref{eq:b} to perform multiplication of a ciphertext by a plaintext, where multiplication is considered as repeated homomorphic addition. 
%It should be noted that while this LWE-based cryptosystem allows for the multiplication of a ciphertext by a plaintext, which can be considered as repeated homomorphic addition. 
However, such an approach %this last operation 
is not recommended as it compromises 
the security and privacy of the closed-loop systems by potential leakage of information from the interception of a plaintext multiplier. %the security of working with just encrypted values by causing leakage of information and reducing the overall security of the system.

\subsection{Infinite Time Horizon Encryption}\label{sec:ITH_enc}

%When we implement the aforementioned ciphertext operations repeatedly in a closed-loop control system with a dynamic controller in the feedback loop, it is prone to {\color{red}the accumulation of error} at every sampling time. Particularly, 
%Performing the previously introduced ciphertext operations, particularly 
%the multiplication operation introduces accumulative damages to the message. % as the error $e$ grows with every ciphertext operation. 
%In order to avoid regular decryption of the data for resetting the error, Kim et al. \cite{Shim_2019} have proposed the following approach that involves the {\color{red} transformation of the controller state matrices defined on reals into those defined on integers}. 
%having to decrypt the data after every operation to reset the error, let us briefly introduce the approach of transforming the controller's state into an integer matrix proposed by Kim et al. \cite{Shim_2019}.
%Let us consider the following discrete-time controllers 

%{\color{green}
%The same meaning for the colors. In my opinion, the problem of growth error and the problem of non-integer state matrix are different problems from each other.
%So, I think the motivation for the integer state matrix should be written in a different way.
%}

{
When we implement the aforementioned ciphertext operations repeatedly in a closed-loop control system with a dynamic controller in the feedback loop, the significand of the encrypted state will tend to accumulate.
A notable exception to this problem is the class of linear dynamic system whose state matrix consists of integers, as observed in
\cite{Integer_Dynamic}.
Let us consider the following discrete-time controllers
}
\begin{equation} 
\label{eq:cont_ITH}
\begin{cases} 
\begin{aligned}
    x(t+1) & = Fx(t)+Gy(t),\\
    u(t) & = Hx(t),
\end{aligned}
\end{cases}
\end{equation}
where, $t\in \mathbb{Z}_+$, $x(t)$ is the controller state variables with the initial values $x_0$, $y(t)$ is the plant's output, $u(t)$ is the controller's output and $F,G,H$ are system's matrices with %$H=\bbm{1 & 0 & \ldots & 0}$ and 
$F$ defined by integers. For the application of HE in the feedback loop, the plant's output $y(t)$ is converted to integers by    
%We ensure the components of the signal $y(t)$ are integers by 
choosing an arbitrary scaling parameter $s_2 \geq 1$ so that %to obtain
\begin{equation} 
 \bar{\bar{y}}(t) := \Bigg\lceil\frac{y(t)}{s_2}\Bigg\rfloor
\label{eq:round_op}
\end{equation}
corresponds to the plant's output in integer form. 
Then, the system is converted into a system over integers by  rounding $G$ using a scaling factor $1/s_1 \geq 1$, such that
{%\color{blue}
\begin{equation} 
\label{eq:cont_ITH_integ}
\begin{cases} 
    \bar{\bar{x}}(t+1) = F\bar{\bar{x}}(t)+\Big\lceil\frac{G}{s_1}\Big\rfloor\bar{\bar{y}}(t),\\
    \bar{\bar u}(t) = \left\lceil\frac{H}{s_3} \right\rfloor \bar{\bar{x}}(t), \\
    \bar{\bar{x}}(0) = \left\lceil\frac{x_0}{s_1s_2}\right\rfloor,
\end{cases}
\end{equation}
where $1/s_3\ge 1$ is the scale factor for the matrix $H$.
Finally, the systems' matrices are encrypted as multipliers as
\begin{equation} 
 \mathbf{F}=\text{Enc2}(F), \  \mathbf{G} =  \text{Enc2}\left(\bigg\lceil\frac{G}{s_1}\bigg\rfloor\right), \  \mathbf{H} = \text{Enc2}\left(\bigg\lceil\frac{H}{s_3}\bigg\rfloor\right).
\label{eq:ITH_vars}
\end{equation}
With the encrypted matrices the dynamic controller becomes
\begin{equation} 
\label{eq:encrypted_controller}
\begin{cases} 
    {\bf{x}}(t+1) = \mathbf{F}\times_\mathcal{C}{\bf{x}}(t)+\mathbf{G}\times_\mathcal{C}{\bf{y}}(t),\\
    {\bf{u}}(t) = \mathbf{H}\times_\mathcal{C}{\bf{x}}(t), \\
    {\bf{x}}(0) =
    \text{Enc}\left( \left\lceil \frac{x_0}{s_1s_2}\right\rfloor \right),
\end{cases}
\end{equation}
where ${\bf x}(t)$ and ${\bf u}(t)$ is the encrypted state and output of the controller, respectively, and
${\bf y}(t)$ is the encrypted plant output, defined as ${\bf y}(t)= \text{Enc}(\bar{\bar y}(t))$.
In case of the first-order system,
the output matrix $H$ can be set as $H=1$, without loss of generality.
Then, the encryption of $H$ as well as the multiplication $\times_\mathcal{C}$ for the output ${\bf{u}}(t)$ becomes dispensable, and we can let
${\bf{u}}(t) = {\bf{x}}(t)$.
}

Regarding the performance error of the encrypted controller,
it is attributable to two factors; error due to quantization (round operation for matrices and signals) and injection of errors during encryption.
We note that the effect of both of them can be arbitrarily small, by appropriate choice of the scale factors $\{s_1,s_2,s_3\}$, and the parameters of the cryptosystem.
See \cite{Shim_2019}, for more details.

%{\color{blue} (Should the time instant $k$ in this subsection be changed to $t$, for consistency with those of (24)?)}

%{\color{green}
%I have just changed the above blue things,
%to accord with the methods of [6], [11], [12].
%Actually I am not sure I am doing right,
%so please check this is fine as it is.
%}

\subsection{Distance-based Distributed Formation Control}

%To control the multi-agent formation we make use of the distributed control laws proposed by De Marina et al. \cite{de2016distributed}. The following paragraphs will outline the mathematics used to control each agent.

For the distributed formation control, we consider the use of well-known distance-based gradient control with estimators as presented in \cite{de2016distributed}. We consider a formation of $n$ mobile robots that move on a plane, e.g., they operates in a 2-dimensional space. %For any given matrix $A$, we denote $\bar{A} = A\otimes I_2$.

The formation of mobile robots/agents can be described using formation graph $\mathcal G = (\mathcal V,\mathcal E)$, where $\mathcal V$ is the set of vertices, each vertex $v\in \mathcal V$ represents an agent, and the set of edges $\mathcal E\subseteq \mathcal V \times \mathcal V$ contains pairs of agents that have to maintain prescribed distance of a given formation shape. We assume that $\mathcal G$ is undirected. %Therefore, graph $G$ with vertex set $V=1,...,n$ and edge set $E\subseteq V \times V$ is the undirected graph of neighbour relationships between agents. 
The set $N_i$ denotes the set of neighbours of agent $i$ given by $N_i := \{ j \in \mathcal V:(i,j) \in \mathcal{E}\}$. For the undirected graph $\mathcal G$, we can define %Then the elements of 
the incidence matrix ${B}=\{b_{ik}\} \in \mathbb{R}^{|\mathcal{V}|\times|\mathcal{E}|}$ by assigning arbitrarily directionality in the graph as follows %can be built with the following conditions
\begin{equation} 
\label{eq:cont_ITH_2}
b_{ik} \triangleq 
\begin{cases} 
\begin{aligned}
    +1\ &\text{if}\ i=\mathcal{E}^{\text{tail}}_{k},\\
    -1\ &\text{if}\ i=\mathcal{E}^{\text{head}}_{k},\\
    0\ &\text{otherwise},\\
\end{aligned}
\end{cases}
\end{equation}
where $\mathcal{E}^{\text{tail}}_k$ and $\mathcal{E}^{\text{head}}_k$ denote the tail and head nodes of edge $\mathcal{E}_k$ respectively. Using the formation graph $\mathcal G$, one can deploy the well-studied gradient-based distributed formation control law where each agent maintains the desired distance with its neighbors in order to form an infinitesimally rigid formation shape \cite{marina2015}. In particular, each robot uses local measurement systems to obtain relative position with respect to the local coordinate system. %When all pair of agents use the same distance information, the desired formation shape is locally exponentially stable. 

%As will be described in detail below, In the following, we will present the use of undirected graph $\mathcal G$  {\color{blue} Bayu will add more.}

Using $\mathcal G$ and the agent positions $p=\bbm{p_1 & p_2 & \ldots &p_n}$, we can define formation shape as follows. Let $\bar{B}=B\otimes I_2$. %The pair $(\mathcal G,p)$ is called the {\it framework}. % is defined by $(G, p)$, where $p$ is the stacked vectors of the agents' positions $p_i$. The sensed 
The measured relative positions defined according to the edge $\mathcal E$ % of each agent 
can now be described by
\begin{equation} 
\label{eq:zzzz}
z=\bar{B}^{T}p,
\end{equation}
where each component $z_k=p_i-p_j$ in $z$ corresponds to the relative position associated with the edge $\mathcal{E}_k=(i,j)$. For a given admissible target shape, we can define a vector of desired inter-agent distances $d^*\in\mathbb{R}^{|\mathcal{E}|}$ defined on $\mathcal E$. The set of all desired equilibrium points is then given by 
\begin{equation}
    \mathcal{D}:=\{p \in \rline^{2n} : \|z_k\| = d_k, k\in \{1,\ldots,\mathcal E\} \}.
    \label{eq:equilibrium_desired}
\end{equation}
%One can check that all points in $\mathcal D$ can be obtained by translation and rotation of a reference position $p^*$ that satisfies the desired distance constraints. 
For detailed analysis on this formation graph, on the characterization of infinitesimally rigid formation and on the standard gradient-based distributed formation control, we refer interested readers to \cite{de2016distributed,AnYuFiHe08,oh2015}.   
%The characterization of infinitesimally

%By defining a vector of desired inter-agent distances $d\in\mathbb{R}^{|\mathcal{E}|}$ such that the error vector $e=z-d$ may be calculated, the gradient based distributed control laws in their closed-loop dynamics are defined as
Assuming that every agent is described by a kinematic point and evolves according to a single integrator, the standard distance-based gradient control law for maintaining the formation is given by 
\begin{equation} 
\label{eq:final_controller_formation}
u=-c_{1}\bar{B}D_z D_{\tilde{z}}e,
\end{equation}
where $\tilde{z}$ is the stacked vector of $||z_k||^{-1}$, $c_1$ is a constant gain, $e=\text{col}_k\{\|z_k\|-d_k^*\}$ is the column vector of the distance error at every edge with $d_k^*$ as the desired distance for the $k$-th edge, $D_z$ and $D_{\bar z}$ represent the block diagonal matrix of $z$ and $\bar z$, respectively. When there is no discrepancy on the distance constraint between pairing agents in all edges and no measurement noise, the control law \eqref{eq:final_controller_formation} ensures local exponential stability to the equilibrium set $\mathcal D$, e.g., the desired formation shape is attained. However, this nice stability property is destroyed when there is a discrepancy/disagreement of the desired distance by agents in an edge. It can happen, for instance, when there is a constant bias in the local range sensor systems. In this paper, we will adopt the solution that is proposed in \cite{marina2015} where a distributed dynamic estimator is deployed to one of the agents in every edge. Particularly, we will adopt the computation of the distributed dynamic estimator in the cloud that is secured by encryption.   %Correspondingly, the closed-loop autonomous systems is described by 
%\begin{equation}
%\begin{aligned}
%&\dot{p}= -\bar{B}D_zD_{\tilde{z}}e,\\
%&\dot{z}= \bar{B}^T\dot{p}, \\
%&\dot{e}= lD_{\tilde{z}}D^{T}_{z}\dot{z}, \\
%\end{aligned}
%\label{eq:closed_loop_de_Marina}
%\end{equation}
%where $e=\text{col}_k\{\|z_k\|-d_k^*\}$ is the column vector of the distance error at every edge with $d_k^*$ be the desired distance for the $k$-th edge, and %, $\dot{p}=u$, with $u$ being the stacked vector of control inputs $u_i\in \mathbb{R}^2$, $\dot{p}$ being the differential of the agents' positions, and 
%$D_z$ represents the block diagonal matrix of $z$. %Therefore, the formation control law is

%The next subsection will elaborate further into how \eqref{eq:final_controller_formation} is modified to allow agents to compensate for mismatches in inter-agent distances, and in turn, obtain the final dynamic control law to be encrypted.

\subsection{Distributed Dynamic Estimators for Mismatch Compensation } %ion of Mismatches Between Agents}

As given in the Introduction, the presence of measurement biases or mismatches $\mu \in \rline^{|\mathcal E|}$ in the distance constraint can induce undesired formation shape and result in group motion as studied in \cite{mou2016}. In these circumstances, de Marina et al. \cite{marina2015} have proposed distributed estimators that can be deployed at a vertex of every edges to robustly compensate these mismatches. %In particular, a dynamic estimator with state $\xi$ and output $\hat\mu$ is proposed and combined with the distance-based distributed formation controller \eqref{eq:final_controller_formation} as follows. 
For each edge $\mathcal{E}$, a local estimator is defined to the agent $\mathcal{E}^{\text{tail}}_{k}$ (called an {\it estimating agent}) and is given by
\begin{align}\label{eq:estimator}
    \begin{split}
        \dot\xi_k &= \kappa (e^{\text{tail}}_k - \hat\mu_k) \\
\hat \mu_k &= \xi_k,
    \end{split}
\end{align}
where the estimator state $\hat{\mu}_k$ aims to compensate for the unknown mismatch or measurement bias $\mu_k$ with respect to the agent $\mathcal{E}_k^{\text{head}}$, and $\kappa>0$ is the estimator gain. Note that $e^{\text{tail}}_k$ and $e^{\text{head}}_k$ is related by $e^{\text{tail}}_k-\mu_k=e^{\text{head}}_k$. Thus when the estimator state $\hat\mu_k(t)\to\mu_k$ as $t\to\infty$, the formation will be defined by $e^{\text{head}}_k$ and the shape will be based on the prescribed distance in $e^{\text{head}}_k$. Using \eqref{eq:final_controller_formation} and \eqref{eq:estimator}, the combined estimator and gradient-based formation control law is given by
%To compensate for possible mismatches in the agent's sensors, the author suggests a dynamic controller capable of removing any potential errors that may cause the formation to not converge. This controller is capable of calculating an estimator $\hat{\mu}$ that converges towards the unknown mismatch $\mu$ in order to cancel it out. The mismatch controller can be combined with the formation controller \eqref{eq:final_controller_formation} to obtain the form of the control law 
\begin{equation} 
\label{eq:estimator_controller}
u=-c_{1}\bar{B}D_z D_{\tilde{z}}e-c_2\bar{B}^{\text{est}}D_{\tilde{z}}\hat{\mu},
\end{equation}
where $c_2$ is a constant gain, $\hat{\mu} \in \mathbb{R}^{|\mathcal{E}|}$ is the stacked column vector of $\hat{\mu}_k$ and $\bar{B}^{\text{est}}=B^{\text{est}}\otimes I_2$ with ${B}^{\text{est}}=\{b_{ik}\} \in \mathbb{R}^{|\mathcal{V}|\times|\mathcal{E}|}$ being a matrix whose elements 
is given by
\begin{equation} 
\label{eq:S_elements}
b_{ik} \triangleq 
\begin{cases} 
\begin{aligned}
    &1\ \text{if}\ i=\mathcal{E}^{\text{tail}}_k,\\
    &0\ \text{if}\ i=\mathcal{E}^{\text{head}}_k,\\
    &0\ \text{otherwise}.\\
\end{aligned}
\end{cases}
\end{equation}
Note that the matrix $\bar{B}^{\text{est}}$ is an indicator matrix for the estimating agent in every edge. %The resulting closed-loop system is given by 
%For each edge $\mathcal{E}$ in graph $G$ we implement a local estimator to the agent $\mathcal{E}^{\text{tail}}_{k}$, called an estimating agent. Therefore, the array $S_1$ contains the choice of estimating agents. To calculate the value of $\hat{\mu}$ for agents in a triangular formation with a cyclical estimation pattern, a closed loop system is used, of the form
%\begin{equation} 
%\begin{aligned}
%\mathbf{P} &:  
%\begin{cases} 
%    \dot{e} = -2e-2(\mu - \hat{\mu}),\\
%    y = e,
%\end{cases}\\
%\mathbf{C} &:    
%\begin{cases} 
%    \dot{\xi} = \kappa(y+\mu - \hat{\mu}), \\
%    \hat{\mu} = \xi,
%\end{cases}
%\label{eq:est_closedloop_lit}
%\end{aligned}
%\end{equation}
%where $\xi$ is the stacked column vector of $\xi_k$, $\mathbf P$ is the error dynamics of the agents and $\mathbf C$ is the dynamic controller. % and the state of the controller.

\section{
Robust Distributed Formation Control with Encryption
%ROBUST DISTRIBUTED FORMATION CONTROL WITH ENCRYPTION
} \label{SEC:designed_system}

In this section, we present the use of an edge/cloud computer to realize the aforementioned distributed formation control with estimator, where the local sensor measurement and distance constraint remain private to the individual agent. In this case, the edge/cloud computer does not act as a centralized computer and is agnostic on the specific knowledge of the agents. The edge computer will compute both the evolution of individual state estimator and the resulting control input in ciphertext. The resulting individual control input is then returned to each agent in ciphertext that can be decrypted using the individual key known only to each agent. This study demonstrates the applicability of HE with infinite-time horizon implementation for the control of multi-agent systems that collaboratively execute a task (maintaining a formation shape, in this case) using only their local information and without sharing information to the neighbours via public edge/cloud computing infrastructure.  %In this case, the edge computer remains agnostic on the heterogeneity of the agents 

%With the underlying mathematics covered, in this section we describe how the dynamic controller \eqref{eq:final_controller_formation} is adapted, and the necessary data are manipulated, to control a triangular ation, as shown in Fig.~\ref{fig:formation_diagram}, with an encrypted controller.
%\begin{figure}[h]
%    \centering
%    \includegraphics[width=0.25\linewidth]{images/Robot Formation.pdf}
%    \caption{Triangular Formation used in simulations with arrows denoting the anti-clockwise estimation pattern.}
%    \label{fig:formation_diagram}
%\end{figure}

\begin{figure*}[!hb] %I put this figure here otherwise it appears on the next page in the Pdf
    \centering
    \includegraphics[width=\linewidth]{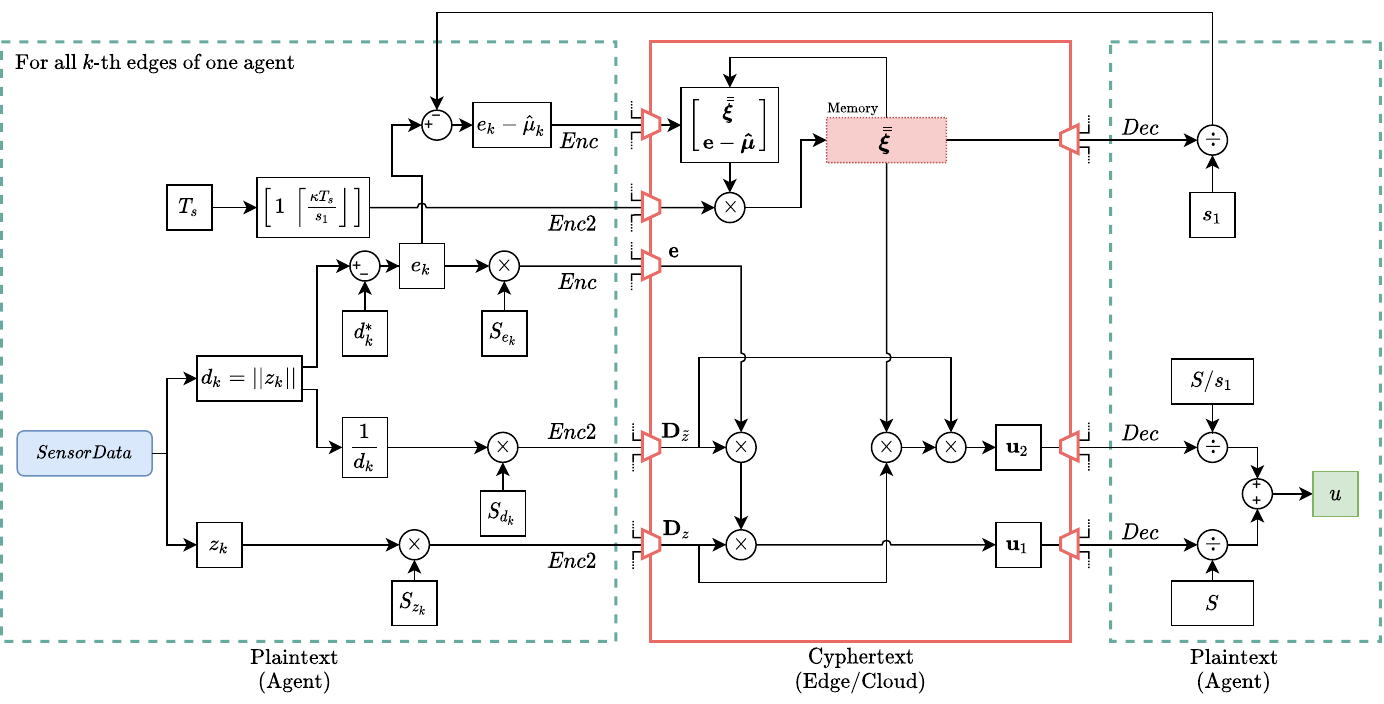}
    \caption{General data flowchart of HE-enabled distributed formation control with estimator which are deployed in the edge/cloud. All variables are described in Section II.C, II.D, III.A and III.B. The isosceles trapezoids represent multiplexers and demultiplexers that combine and split data from and to all edges, respectively. For brevity, the operations in the edge/cloud uses the concatenated data while in practice, they can be computed ``locally'' in the edge/cloud per agent. The input data to the flowchart is the on-board sensor measurement and the output data is the control input $u$ to be implemented in each robotic agent. } %The variable $z_k$ is the relative distance between agents at the $k$-th agent. $d^*$ is the desired inter-agent distance, and $u$ is the speed of the agent. Finally, $S_\ell$ is the scaling applied to variable $\ell$, and $S$ is the multiplication of all $S_\ell$.}
    \label{fig:dataflow}
\end{figure*}

\subsection{Scaled Logarithmic Quantization for Scaling Data}

In the application of HE, one crucial element to communicate real-time data between the agent and the edge/cloud computer is the use of scaled quantization.   %from the agent's sensors. 
Given a real-time value of a variable $\nu_k\in\mathbb{R}$, which is available to an agent in the $k$-th edge (e.g., the distance $d_k$, the elements in relative position vector $z_k$, etc), it will be scaled (up or down) and rounded so that it can be encrypted based on the allocated plaintext space. %we need to transform it to integer so that it can be encrypted and transported to the edge computer. 
As commonly done, we can perform the scaled quantization to the variable $\nu_k$ %scale and round it by %doing the operation 
%it is possible to obtain 
by $\bar{\bar{\nu}}_k= \lceil S_{\nu_k} \nu_k\rfloor \in \mathbb{Z}$ where the scaling factor $S_{\nu_k}>1$ must be chosen appropriately. The same scaling factor will be used again to rescale the decrypted control input from the edge/cloud computer. %which is parametrized by $\ell \in \mathbb{Z}$. %for variable $k$. 

In general, there are two issues in the use of static scaling factor $S_{\nu_k}$. Firstly, the scaled quantization is known to % This procedure will 
introduce scaled quantization error that will affect the stability of the closed-loop system. In particular, when there is a memory element in the feedback loop, the memory state can accummulate such quantization error that can deteriorate the closed-loop systems performance and hence it requires regular resetting of controller state. Secondly, as the use of static scaling factor is the same as using uniform quantizer, we can only guarantee practical stability where steady-state error will occur due to the rounding precision. We refer to \cite{depersis2012} on general practical stability analysis for passive systems using such uniform quantization, which represents also the mobile robot dynamics with gradient-based control law.    

%propagate back to the system affect the resolution of data used to calculate the input velocity for an agent. 
Inspired by the use of logarithmic quantizer, which can guarantee asymptotic stability of closed-loop systems \cite{fu2005}, we consider in this work the use of scaled logarithmic quantizer where the scaling factor $S_{\nu_k}$ changes in a logarithmic fashion. 
Intuitively, we want to maximize $S_{\nu_k}$ to ensure that we maintain the precision and stability of the system. However, as defined in Section \ref{SEC:FHE_pre}, the upper-bound condition on the message $|m| < p/2$ limits our ability to scale the data arbitrarily. %to the size of available plaintext space $p$. 

%In this subsection we elaborate on how the scaling factors $S_{\nu_k}$ can be chosen for every encrypted variable to ensure that the controller realized on the edge computer can run with infinite time horizon and converge to the desired formation. %has enough information to reach convergence.

\begin{figure}[h]
    \centering
    \includegraphics[width=\linewidth]{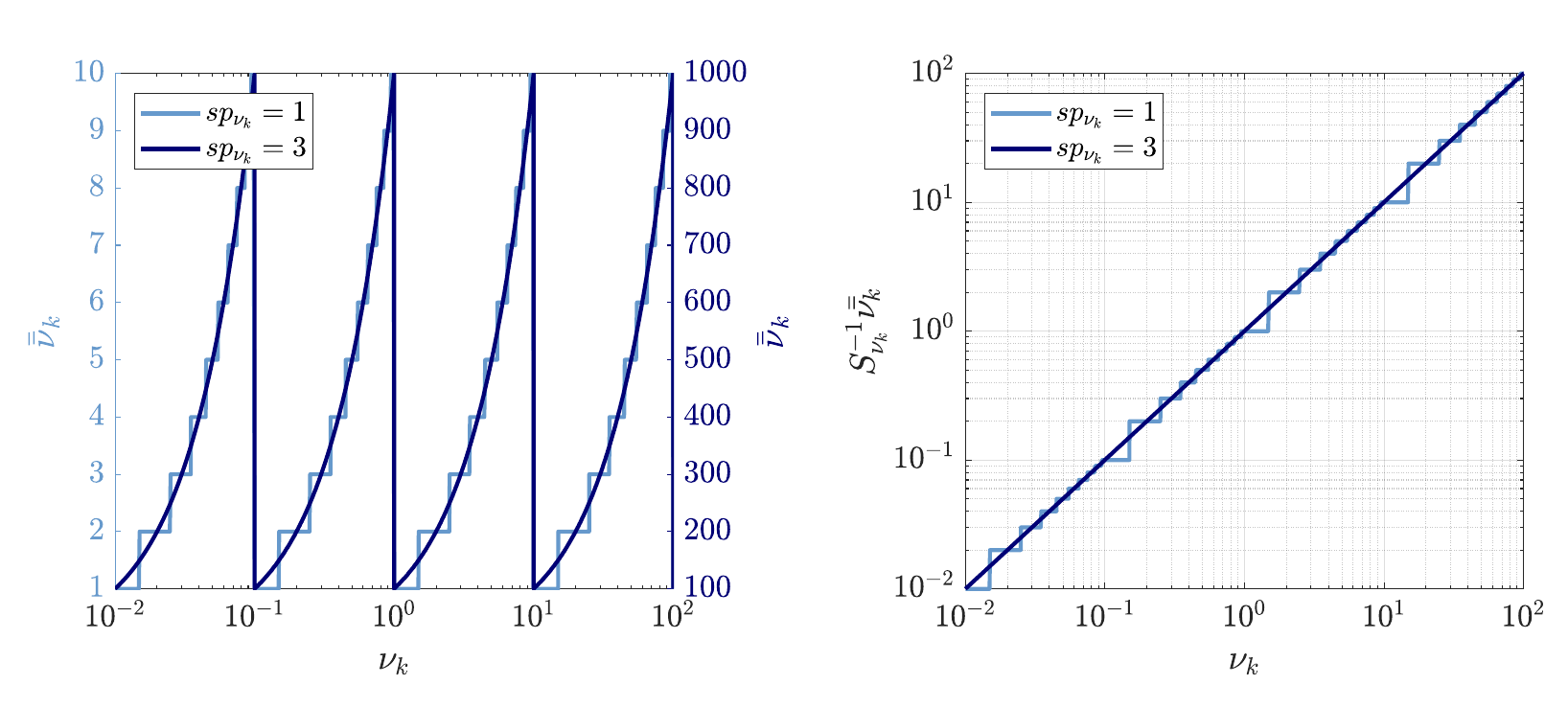}
    \caption{Left: The plot of $\bar{\bar{\nu}}_k= \lceil S_{\nu_k} \nu_k\rfloor$ with $S_{\nu_k}$ be as in \eqref{eq:significant figures} and with $sp_{\nu_k}=1$ (in light blue) and $sp_{\nu_k}=3$ (in dark blue). The abscissa is $\nu_k$ in log-scale and ordinate is $\bar{\bar{\nu}}_k$; Right: The plot of $S_{\nu_k}^{-1}\bar{\bar{\nu}}_k$ in log-scale with $sp_{\nu_k}=1$ (in light blue) and $sp_{\nu_k}=3$ (in dark blue). }
    \label{fig:log_scal}
\end{figure}

The logarithmic design of $S_{\nu_k}$ is based on prioritizing the significant figure of $\nu_k$ than its decimal values. Accordingly, for each variable of interest $\nu_k$, we define a desired significant figures $sp_{\nu_k}$ and the scaling factor is given by   
%we must note that in the scaling process we prioritize significant figures rather than decimal values. Hence, by defining a desired significant figures $sp_\ell$, % for variable $k$, 
%we define $S_\ell$ by
\begin{equation} \label{eq:significant figures}
%\begin{aligned}
S_{\nu_k} = 10^{sp_{\nu_k}-\lfloor \log_{10}(|\nu_k|)\rfloor -1}.
%\end{aligned}
\end{equation}
This allows us to send the information in the data up to a prescribed significant number. 

Using the logarithmic scaling factor as above, we have that changes in the magnitude of the measures as the agents move closer or further apart do not affect the maximum magnitude of the information sent. Figure \ref{fig:log_scal} shows the plot of the scaled logarithmic quantization for different value of  $sp_{\nu_k}$. It shows that the scaled logarithmic quantizer can approximate an identity operator well in contrast to the use of uniform quantizer that becomes zero below a certain threshold value. As illustrated in this figure as well, the choice of significant figure $sp_{\nu_k}$ may cause the message to be larger than the allowable plaintext space, i.e., $|\bar{\bar{\nu}}_k| > p/2$. Therefore, we will now define the next step to determine the bounds for the plaintext space $p$. %The next step to define possible values for $S_k$ is to determine the boundaries of $p$. 
Note that the minimum required plaintext space can be calculated based on the desired significant figures and amount of ciphertext operations to be carried out. Based on a number of multiplied variables $m$ with different $sp_i$, $i\in\{1,...,m\}$, and based on the number of added variables $n$ with different $sp_j$, $j\in\{1,...,n\}$, we can have the following conservative lower bound of $p$%an approximated ideal lower-boundary of $p$ is  
\begin{equation} 
p > \prod_{i=0}^{m}{10^{2sp_i}}+2\sum_{j=0}^{n}{10^{sp_j}}.
\label{eq:scaling_gen}
\end{equation}
On the other hand, the upper bound of $p$ can generally be determined by the precision of the data format used to perform the calculations. A 64-bit signed integer would require $p<2^{63}-1$. 
Although this maximum is enough to contain the desired messages, there is a caveat of defining a large $p$ in the application process. That is, the magnitude of $p$ also defines the upper boundary $Lp$ of the elements of a ciphertext. Therefore, the resulting values of one multiplication operation will have a maximum of $(Lp)^2$. Using signed integers this quickly results in data overflow. Hence, at the expense of memory and performance, we avoid this limit by making use of an arbitrary-precision integer which is provided, for instance, by the programming language Python. %Although less efficient, we found the compromise necessary to choose a $p$ large enough that allowed the formation to converge.

\subsection{Encrypting Estimator Dynamics}

As we implement the distributed estimators with its own state variable in the edge/cloud, we encrypt them as follows. First we consider the following discrete-time version of \eqref{eq:estimator} 
%\begin{subequations}\label{eq:est_to_disc_1}
\begin{align}\label{eq:est_to_disc_1}
    \begin{split}
        \xi(t+1) & = \xi(t) +  T_s\kappa(e^{\text{tail}}(t) - \hat{\mu}(t))\\
        \hat{\mu}(t) & = \xi_k(t)
    \end{split}
\end{align}
where $t\in \mathbb{Z}_+$ is the discrete-time, and $T_s>0$ is the sampling time. 
%The first step to encrypt \eqref{eq:estimator_controller}, using the methodology covered in Section \ref{sec:ITH_enc}, is to discretize the controller used to obtain the estimator $\hat{\mu}$. Considering the plant and controller \eqref{eq:est_closedloop_lit} in their general compact form, we discretize the controller considering a sampling time $T_s<1$ such that
%\begin{equation}
%\label{eq:AA}
%    \dot{\xi}(t) = \frac{{\xi}(k+1) -{\xi(k)}}{T_s}.
%\end{equation}
%Substituting \eqref{eq:AA} into the controller and taking $t = k$, we obtain
%\begin{equation}
%\label{eq:est_to_disc_1}
%    \frac{{\xi}(k+1) -{\xi(k)}}{T_s} = \kappa(y+\mu - \hat{\mu}).
%\end{equation}
%Then, rearranging \eqref{eq:est_to_disc_1}, we obtain the discretized controller in matrix form
%\begin{equation} 
%\label{eq:final_mat_form}
%\begin{array}{rl}
%\xi_k(t+1) & = 
%\begin{bmatrix}
%I & \kappa T_s  
%\end{bmatrix}
%\begin{bmatrix}
%\xi_k(t)\\
%(e_k^{\text{tail}}(t) - \hat{\mu}_k(t))
%\end{bmatrix} \\
%\hat \mu_k(t) & = \xi_k(t)
%\end{array}
%\end{equation}
%Or, equivalently, 
%\begin{equation} 
%\label{eq:final_mat_form}
%\begin{array}{rl}
%\xi_k(t+1) & = 
%(1-T_s\kappa) \xi_k(t) +  T_s\kappa e_k^{\text{tail}}(t) \\
%\hat \mu_k(t) & = \xi_k(t).
%\end{array}
%\end{equation}
Correspondingly, with a scale factor $1/ s_1 \ge 1$,
we convert the system \eqref{eq:est_to_disc_1} as
\begin{equation}\label{eq:final_mat_form}
\bar{\bar{\xi}}_k(t+1)  = 
\bar{\bar{\xi}}_k(t) +  \left\lceil\frac{T_s\kappa}{s_1}\right\rfloor \Big(e_k^{\text{tail}}(t)-\hat\mu_k(t)\Big),
\end{equation}
in which, the error $\| \xi_k(t) - s_1\bar{\bar \xi}_k(t) \|$ tends to zero as $s_1$ tends to zero,
so that the output $\hat \mu_k(t)$ can be obtained by $\hat \mu_k(t)  = s_1\bar{\bar{\xi}}_k(t)$.
%we apply scaling to \eqref{eq:est_to_disc_1} by
%$s_1 \leq 1$, e.g.,   $\bar{\bar{\xi}}_k=\frac{\xi_k}{s_1}$ so that $\frac{\kappa T_s}{s_1}\in \mathbb{Z}$, which gives us
As the equation \eqref{eq:final_mat_form} uses only integer coefficients, it can directly be implemented in the edge/cloud environment using HE while the data of $e_k^\text{tail}(t)-\hat\mu_k(t)$ is transmitted by the agent as integers via standard scaled quantizer. Hence, the encrypted estimator state variable will evolve as an integer in the ciphertext and, when it is returned back to the agent, its value is decrypted and scaled back by the same scaling constant $s_1$.  
%\begin{equation} 
%\label{eq:final_mat_form}
%\frac{\xi(k+1)}{s_1}=\bar{\bar{\xi}}(k+1)= 
%\begin{bmatrix}
%I & \frac{\kappa T_s}{s_1}  
%\end{bmatrix}
%\begin{bmatrix}
%\bar{\bar{\xi}}(k)\\
%y+\mu - \hat{\mu}
%\end{bmatrix}
%\end{equation}

As the estimator state always evolves in the ciphertext, it can be implemented with an infinite time horizon. On the other hand, it has been shown in \cite{marina2015} that the closed-loop system (without encryption) is locally exponentially stable, so that it is also locally input-to-state stable. Therefore, the error due to the use of scaled logarithmic quantizer will lead to practical stability, e.g., steady-state error can occur which is close to zero as a logarithmic scale is used. % In this case, the resulting error will practically be close to zero as logarithmic scaled quantizer is used. 

In order to secure the privacy of the coefficients of the estimators, we apply encryption Enc2 to the coefficients $1$ and {$\lceil\frac{T_s\kappa}{s_1}\rfloor$} which can be computed offline and stored in the edge/cloud. 
%Using \eqref{eq:final_mat_form}, and due to $I$ being an integer state matrix, the controller can be encrypted such that the estimator $\xi(k+1)$ may remain encrypted for an infinite time horizon. Hence, it must be noted that to adapt the equation into the applied digital system, the matrix $[I, \kappa T_s/s_1]$ must be encrypted as the multiplicand by using the $\text{Enc2}(\cdot)$ function to be able to perform the ciphertext multiplication. 

\subsection{Dataflow Diagram}
Based on the description in the previous subsections, we can now present the architecture of the HE-enabled formation control with mismatch estimator. Figure \ref{fig:dataflow} shows the data flowchart from the edges of an agent to the edge/cloud where the distributed formation control computation is performed and returns the formation control input according to \eqref{eq:estimator_controller} implemented in the ciphertext. As shown in this figure, the scaled logarithmic quantization is applied to the distance-based formation control part (e.g., \eqref{eq:final_controller_formation}), while the estimator part (the last term in \eqref{eq:estimator_controller} and \eqref{eq:estimator}) uses standard scaled quantization with fixed $S_{\nu_k}$ prior to the encryption. %The encrypted control input is deciphered and re-scaled by the agent before it is implemented in the robot. 
Here, the decrypted $\hat\mu(t+1)$ is used to compute $e_k(t+1)-\hat\mu(t+1)$ for the next sampling time. 
%Given the difference between $\text{Enc}()$ and $\text{Enc2}()$ covered in Section~\ref{SEC:FHE_pre}, each variable must be encrypted considering its expected purpose in the ciphertext operations. Fig.~\ref{fig:dataflow} shows our design for how each variable is manipulated and encrypted to perform the controller operations for one agent in a triangular formation in ciphertext form.

\section{Simulation Results} \label{SEC:simu_set}
\subsection{Simulation Setup}

\begin{figure}[h]
    \centering
    \includegraphics[width=0.25\linewidth]{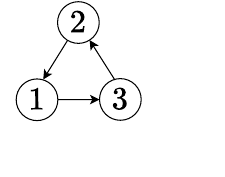}
   \caption{The formation graph used in simulations where the three agents are represented by the three nodes. The direction of the graph indicates the estimating agent for each edge, i.e. the node at the tail of each arrow.}
    \label{fig:formation_diagram}
\end{figure}

In this section, we show and evaluate the performance and robustness of the proposed HE-enabled distance-based formation control with estimators. % encrypted formation controller, in this section we present some results obtained from various simulations. For this purpose 
For simplicity, we consider the formation of three agents forming a triangle. Figure \ref{fig:formation_diagram} shows the formation graph where estimating agents are shown at the tail of each edge. 

%\begin{figure}[h]
%    \centering
%    \includegraphics[width=\linewidth]{images/Nodediagram.pdf}
%    \caption{ROS Node-Topic map to calculate the encrypted controller of one agent. Orange  represents nodes, while blue represents a topic, i.e. message sent between nodes.\textcolor{blue}{Mariano: I am thinking this image may not be necessary, and we can save the space by deleting it...}}
%    \label{fig:Node_diagram}
%\end{figure}

All simulations were carried out on a laptop running ROS \cite{ROS} Melodic in Ubuntu 18.04 with an i7-4700MQ CPU, GT 740M GPU and 8GB of RAM. For the simulation setup in the Gazebo environment, we use LIDAR as the on-board distance sensor in each robot. The ROS robot model is based on the Nexus$^\text{\textregistered}$ mobile robot equipped with four 100mm Mecanum$^\text{\textregistered}$ wheels. The control input will be the longitudinal and lateral velocity of the robot. For all agents, the desired distances for the distance-based formation control are set to $d_k^*=0.8$, with a sampling time $T_s=0.1[s]$, and initial estimator state $\bar{\bar{\xi}}_k = 0$. %To measure the distance between neighbouring agents a LIDAR sensor is used on each one. This tool allows for the detection of other agents by using the laser in the sensor as it revolves. However, the sensor is not able to differentiate between agents and obstacles, therefore, the robots are assumed to be in an empty room.
Unless stated otherwise, the encryption within the simulations was carried out with a key length of $N=10$, an injected error of $e=100$, and an available plaintext space of $p=10^{13}$.

%\section{SIMULATION RESULTS}

\subsection{Simulation Result}

\begin{figure}[ht]
    \centering
    \includegraphics[width=\linewidth]{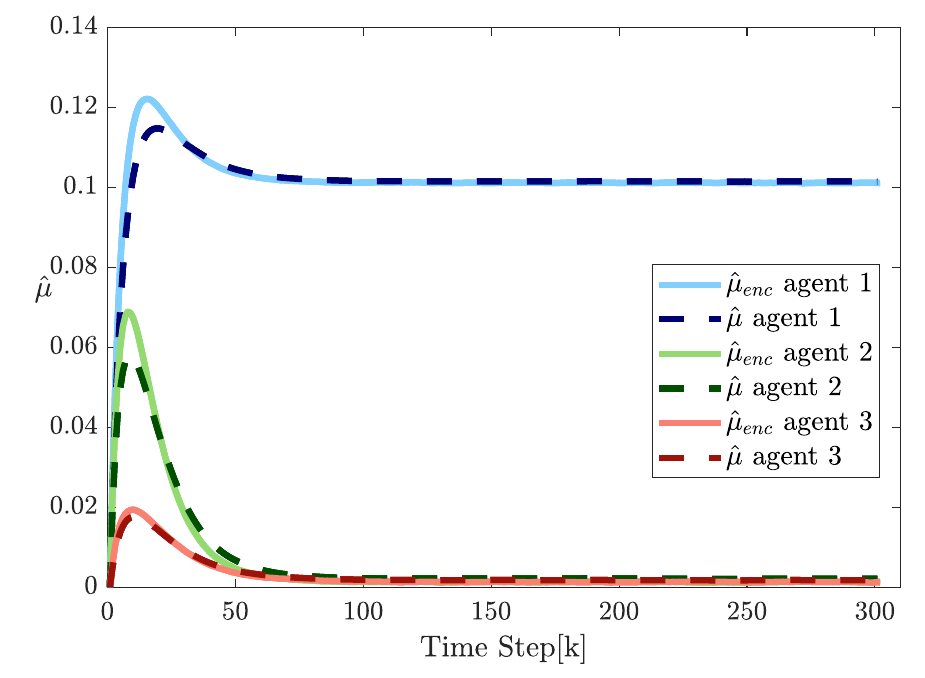}
    \caption{Simulation result of the closed-loop system with and without encryption in the formation control. The plot shows the %of \eqref{eq:estimator_controller} comparing 
    settling performance of $\hat{\mu}_{enc}$ (encrypted) against $\hat{\mu}$ (non-encrypted) %for a formation of robots as shown in Fig. \ref{fig:formation_diagram}, 
    where agent 1 is assigned a mismatch of $\mu_k = 0.1$ with respect to its neighbor. }%The data generated supports that, using the proposed system, $\hat{\mu}$ may remain encrypted without significant effects to its settling point.}
    \label{fig:Matlab_mu}
\end{figure}

For comparing the performance of the closed-loop systems with and without encryption with infinite-time horizon, we performed a simulation of the closed-loop system for both scenarios where agent 1 has an initial distance error of $e_k=0.1$ and the same distance mismatch of $\mu_k=0.1$ with respect to the other two agents. Figure \ref{fig:Matlab_mu} shows the performance of the HE-enabled formation control is close to that of the original one. In particular, the encrypted estimator states are able to converge to the correct ones with similar convergence rate as the original one. More importantly, in this simulation, the evolution of estimator states remains encrypted throughout simulation time without any problem. 

%To ensure the encryption of the estimator $\hat{\mu}$ is stable for an infinite time horizon, we carried out an initial simulation of \eqref{eq:estimator_controller} independent of sensor physics. In this scenario, agent 1 is assigned a distance error $e=0.1$ from other agents, and a mismatch $\mu=0.1$. Given these conditions, the results shown in Fig.~\ref{fig:Matlab_mu} display the estimator's capability to converge and maintain its value for 300 time steps. That is, the encrypted state was updated 300 times in its ciphertext form without any considerable negative effects on the settling point.

%\subsection{Encrypted Control of Robot Formation}

%To ensure the formation is able to converge, we set the agents at a distance from each other of 1 [m], with a goal distance of 0.8 [m].

\subsection{Robustness Analysis}

One important aspect to the security of a homomorphically encrypted system is the role of key length $N$. In this regard, we perform robustness analysis of the formation convergence via Monte Carlo analysis for multiple values of $N$. %present the results of a formation's convergence over multiple iterations with various values of N. 
For brevity, we only present the convergence of one agent's distance to one of its neighbours. The similar behaviour is also exhibited by the other agents. 

\begin{figure}[ht]
    \centering
    \includegraphics[width=\linewidth]{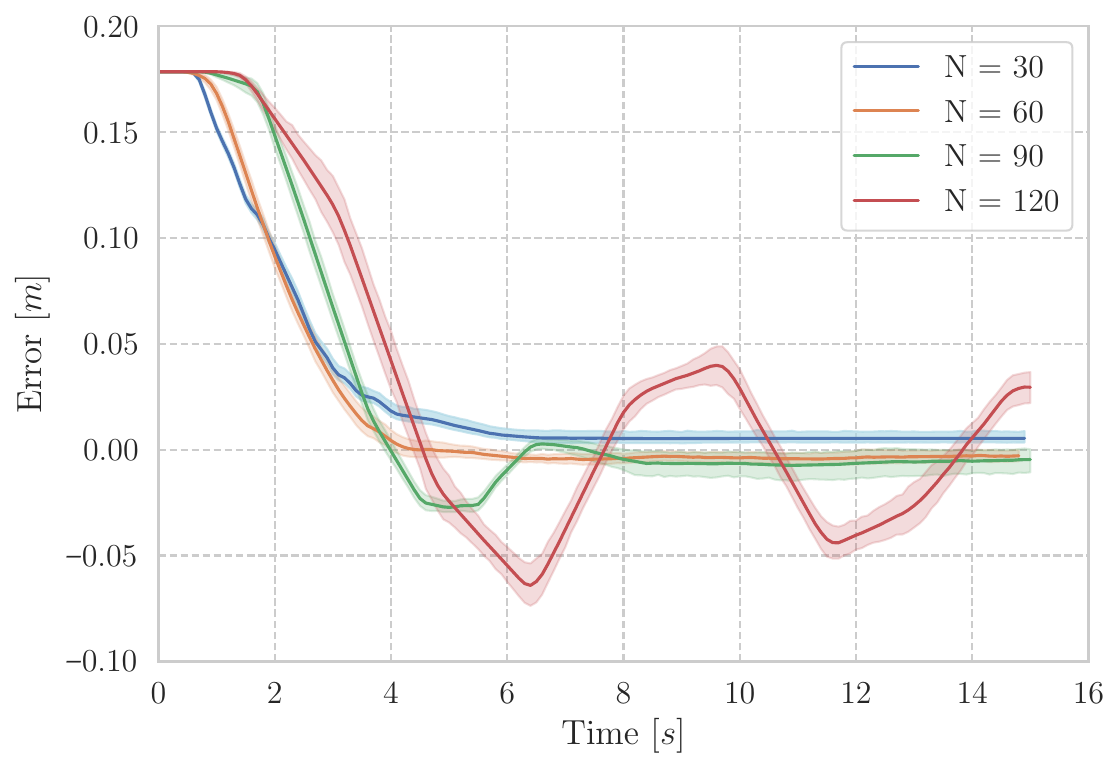}
    \caption{Effect of key length $N$ on an agent's transient behaviour of the error in distance with respect to a neighbour. %over time relative to a neighbouring agent. 
    A 95\% confidence interval indicates the distribution of convergence over 50 simulations per value of $N$.} % with a desired distance of 0.8 [m]. 
    %The magnitude of N causes the encryption process to take longer, therefore, the larger overshoots occur due to the slower update time of the robot's velocity.}
    \label{fig:N10_35}
\end{figure}

\begin{figure}[ht]
    \centering
    \includegraphics[width=\linewidth]{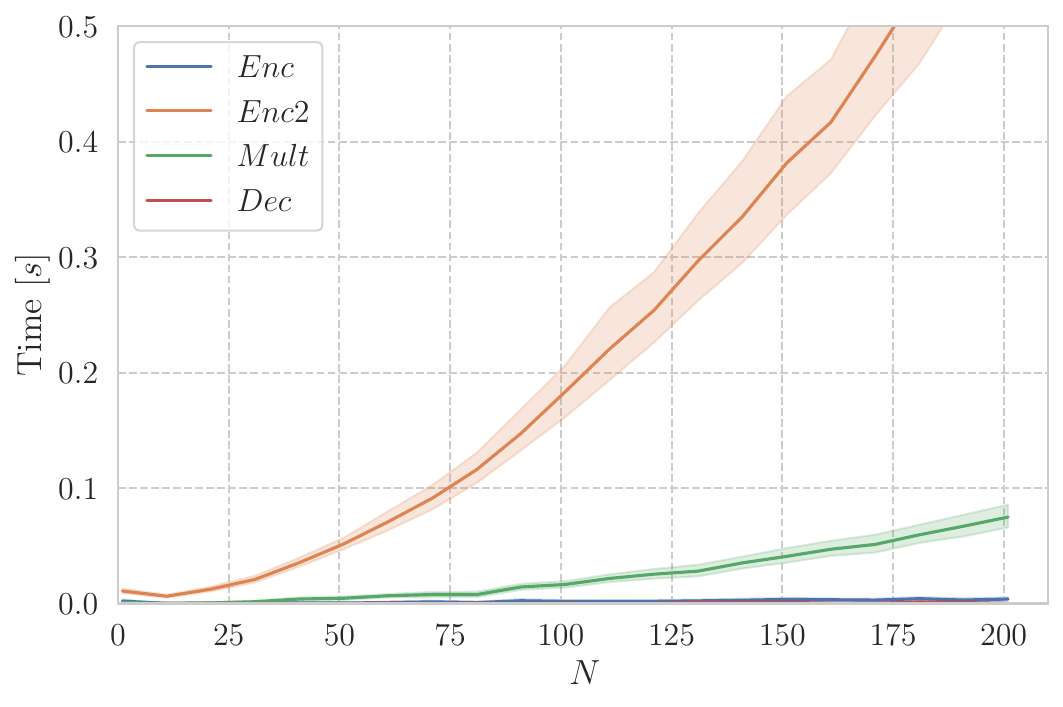}
    \caption{Effect of $N$ on run-time execution of encryption ($Enc$, $Enc2$) and decryption ($Dec$) of an integer, as well as multiplication ($Mult$) of two encrypted values. A 95\% confidence interval indicates the distribution of run-time over 50 simulations per operation type.}% Longer times result in slower convergence of formation. (Violin plot)}
    \label{fig:Time_enc}
\end{figure}

%To ensure the system's stability we ran the same simulation for each value of N fifty times.
We consider four different values of $N=30, 60, 90$ and $120$ and for each value of $N$, we run fifty simulations with initial distance of $d_k=1$ and constant mismatch of $\mu_k=0.1$ in agent $1$. In this simulation, the source of uncertainties is on the computational resources for the encryption and for control computation that introduce overhead and asynchronous simulation time in each simulated robot. Figure \ref{fig:N10_35} shows the transient behaviour for all $N$ with 95\% confidence interval indicating the aforementioned variability between simulations. This simulation shows degradation in the convergence of formation for larger values of $N$, which is mainly due to the update time of the agent's velocity caused by the encryption. In Figure \ref{fig:Time_enc}, we plot the measured run-time of the executed operations on one integer for increasing $N$ and %By measuring the time taken to perform the encryption, in Fig.~\ref{fig:Time_enc} 
we can see that the increase in key length dramatically affects the time taken to encrypt a value using Enc2, which directly impacts the time between control inputs to an agent. 
%In every iteration all agents are assigned a desired distance $d^*=0.8$, with an initial distance of $d_k=1$, with a constant mismatch $\mu=0.1$ in agent 1. The results are shown in Fig.~\ref{fig:N10_35} with a 95\% confidence interval indicating the variability between simulations.

%\begin{figure}[h]
%    \centering
%    \includegraphics[width=0.9\linewidth]{images/Bitsecurity.pdf}
%    \caption{Considering the rest of the variables as constant but increasing $N$ the bit-security of HE is displayed, where a value of more than 1500 would be required to surpass current standards \cite{Bit-security}. \textcolor{blue}{Mariano: (Question for Junsoo) I used a figure like this in my thesis to illustrate the security provided by HE compared to existing methods like AES, but delving more into it I learnt that bit-security is very dependent on the attack. Therefore, I am afraid this may not be accurate enough for this paper.}}
%    \label{fig:bitsecurity}
%\end{figure}

\section{Conclusions and Future Works} \label{SEC:conclusions}

In this work, we presented an HE-enabled distributed formation control with estimators. We proposed the use of a scaled logarithmic quantizer in combination with the transformation of the estimator to operate over integers. This allows us to secure the privacy of the controller in the edge/cloud computer where the estimator states evolve completely in ciphertext (i.e., it is always  encrypted) with infinite-time horizon. Simulation results are presented that show the efficacy and robustness of the proposed architecture. % a system design for the encrypted control of a dynamic formation controller. By applying homomorphic encryption theory on a distributed formation controller we contribute an encrypted capable of a technique to scale data such that... 

%Subsequently, by simulating the encrypted system we show the formation maintains its ability to converge to a desired inter-agent distance 

%The scheme presented for the encrypted controller of a robot formation allows us to simulate a homomorphically encrypted formation. Furthermore, by transforming the estimating controller to operate over integers, it is made possible to maintain a variable constantly encrypted. However, 
There are some limitations that affect the design of the encrypted controller and the simulations that should be considered in future approaches. The limited computational power has affected the convergence time when larger $N$ (improved security) is used. One can deploy high-computing infrastructure to ensure that the encryption process does not become the bottleneck in the edge/cloud so that the convergence time is not affected.  

The use of single integrator as the estimator is to compensate for constant mismatches in the distance constraints. When there are (periodic) disturbance/reference signals that are generated by known exosystems, we can also deploy dynamic internal models in the edge/cloud.
%A considerable weak point in the simulation setup shown in Section \ref{SEC:simu_set} is the limited computing power used to perform the encryption. %\footnote{All simulations shown in this paper were carried out on a laptop running ROS \cite{ROS} Kinetic in Ubuntu 16.04 with an i7-4700MQ CPU, GT 740M GPU and 8GB of RAM}. It could be possible to obtain faster encryption by using a more powerful computer, and by extension, the potential to obtain faster convergence with larger values of N for improved bit security.

In order to compute the large integers generated in the encryption process we made use of arbitrary precision arithmetics offered natively in Python. Also termed ``multiple precision'', this challenge is reoccurring in the implementation of cryptography \cite{st2006bignum}. Although less efficient than using fixed-precision data types, we found it fast enough to perform the needed operations for this proof of concept. Further research is necessary to analyse the concrete impact on the encryption time, and settling of the formation. 

Finally, we provide some remarks in dealing with the computational complexity in HE-enabled distributed control systems:
\begin{itemize}
    \item {\it Encryption of static parameters off-line}.
    The encryption of the matrix parameters as static multipliers
    can be performed off-line, during initialization.
    Thus,
    despite the applied encryption algorithm for the matrices requiring a relative large amount of computational resources,
    it would not be a burden for
    real-time control operation.
    \item {\it The use of non-private data as unencrypted}. 
    For example,
    by making public the use of the integral controller in the proposed method, the associated state 
    matrices that do not contain any private information can be kept unencrypted. 
%    The less portion of data needed to be encrypted, the less computational complexity will be required.
\end{itemize}

%\textbf{(WIP)}
% The design of the encrypted system could be improved by:
%\begin{itemize}
    %\item Junsoo: Identity matrix could remain public to reduce computational burden
    %\item Junsoo: element $Ts/s_1$ could be encrypted a priori during initialization to reduce computing burden
%    \item computing the final addition encrypted. Currently it cannot be calculated as a ciphertext due to the difference in scaling factors. \eqref{eq:final_controller_formation}
%\end{itemize}

%The encrypted system could be applied to a real NCS. The limitations of the real world, such as network lag or sensor accuracy, could provide further insight into the application challenges of HE. 

%%%%% -12
\balance %Using this package because the following line is not working
\addtolength{\textheight}{-5cm}% This command serves to balance the column lengths

%%%%%%%%%%%%%%%%%%%%%%%%%%%%%%%%%%%%%%%%%%%%%%%%%%%%%%%%%%%%%%%%%%%%%%%%%%%%%%%%

%%%%%%%%%%%%%%%%%%%%%%%%%%%%%%%%%%%%%%%%%%%%%%%%%%%%%%%%%%%%%%%%%%%%%%%%%%%%%%%%

%%%%%%%%%%%%%%%%%%%%%%%%%%%%%%%%%%%%%%%%%%%%%%%%%%%%%%%%%%%%%%%%%%%%%%%%%%%%%%%%
%\section*{APPENDIX}

\section*{Acknowledgement}
We  would  like  to  thank  prof. Hyungbo Shim from Seoul National University for the initial discussion and problem formulation setup. %his extensive contributions to the homomorphic encrypted control theory we made use of in this work.

%%%%%%%%%%%%%%%%%%%%%  BIBLIOGRAPHY  %%%%%%%%%%%%%%%%%%%%%

%\cite{IEEEexample:article_typical}

\bibliographystyle{IEEEtran}
%\bibliography{IEEEabrv,IEEEexample}
\bibliography{IEEEabrv,bibliography}

%\begin{thebibliography}{99}
%\bibitem{c1} G. O. Young, ÒSynthetic structure of industrial plastics (Book style with paper title and editor),Ó 	in Plastics, 2nd ed. vol. 3, J. Peters, Ed.  New York: McGraw-Hill, 1964, pp. 15Ð64.
%\bibitem{c2} W.-K. Chen, Linear Networks and Systems (Book style).	Belmont, CA: Wadsworth, 1993, pp. 123Ð135.
%\bibitem{c3} H. Poor, An Introduction to Signal Detection and Estimation.   New York: Springer-Verlag, 1985, ch. 4.
%\bibitem{c4} B. Smith, ÒAn approach to graphs of linear forms (Unpublished work style),Ó unpublished.
%\end{thebibliography}

\newpage
%\onecolumn
%\include{Review_ToDo}
%when deleting these three lines change back "addtolegth" to -12, and delete package "ulem"

\end{document}